\documentclass[aps,prl,twocolumn,superscriptaddress,amsmath,amssymb,10pt]{revtex4}
\usepackage{graphicx}
\usepackage{amsmath}
\usepackage{latexsym}
\usepackage{dcolumn}
\usepackage{bm}
\usepackage{float}
\usepackage{graphicx,here}

\emergencystretch=\maxdimen
\begin{document}

\title{Superconductivity and Fermi Surface Nesting in the Candidate Dirac Semimetal NbC}

\affiliation{Institute of Physics, Chinese Academy of Sciences, Beijing 100190, China}
\affiliation{Beijing Academy of Quantum Information Sciences, Beijing 100193, China}
\affiliation{Hiroshima Synchrotron Radiation Center, Hiroshima University, 2-313 Kagamiyama, Higashi-Hiroshima 739-0046, Japan}
\affiliation{School of Physical Sciences, University of Chinese Academy of Sciences, Beijing 100049, China}
\affiliation{Center of Materials Science and Optoelectronics Engineering, University of Chinese Academy of Sciences, Beijing 100049, China}
\affiliation{Songshan Lake Materials Laboratory, Dongguan, Guangdong 523808, China}

\author{Dayu Yan}
\affiliation{Institute of Physics, Chinese Academy of Sciences, Beijing 100190, China}
\affiliation{School of Physical Sciences, University of Chinese Academy of Sciences, Beijing 100049, China}
\affiliation{Center of Materials Science and Optoelectronics Engineering, University of Chinese Academy of Sciences, Beijing 100049, China}
\author{Daiyu Geng}
\affiliation{Institute of Physics, Chinese Academy of Sciences, Beijing 100190, China}
\affiliation{School of Physical Sciences, University of Chinese Academy of Sciences, Beijing 100049, China}
\author{Qiang Gao}
\affiliation{Institute of Physics, Chinese Academy of Sciences, Beijing 100190, China}
\affiliation{School of Physical Sciences, University of Chinese Academy of Sciences, Beijing 100049, China}
\author{Zhihai Cui}
\affiliation{Institute of Physics, Chinese Academy of Sciences, Beijing 100190, China}
\affiliation{School of Physical Sciences, University of Chinese Academy of Sciences, Beijing 100049, China}
\author{Changjiang Yi}
\affiliation{Institute of Physics, Chinese Academy of Sciences, Beijing 100190, China}
\affiliation{School of Physical Sciences, University of Chinese Academy of Sciences, Beijing 100049, China}
\affiliation{Center of Materials Science and Optoelectronics Engineering, University of Chinese Academy of Sciences, Beijing 100049, China}
\author{Ya Feng}
\affiliation{Beijing Academy of Quantum Information Sciences, Beijing 100193, China}
\author{Chunyao Song}
\affiliation{Institute of Physics, Chinese Academy of Sciences, Beijing 100190, China}
\affiliation{School of Physical Sciences, University of Chinese Academy of Sciences, Beijing 100049, China}
\author{Hailan Luo}
\affiliation{Institute of Physics, Chinese Academy of Sciences, Beijing 100190, China}
\affiliation{School of Physical Sciences, University of Chinese Academy of Sciences, Beijing 100049, China}
\author{Meng Yang}
\affiliation{Institute of Physics, Chinese Academy of Sciences, Beijing 100190, China}
\affiliation{School of Physical Sciences, University of Chinese Academy of Sciences, Beijing 100049, China}
\affiliation{Center of Materials Science and Optoelectronics Engineering, University of Chinese Academy of Sciences, Beijing 100049, China}
\author{Masashi Arita}
\affiliation{Hiroshima Synchrotron Radiation Center, Hiroshima University, 2-313 Kagamiyama, Higashi-Hiroshima 739-0046, Japan}
\author{Shiv Kumar}
\affiliation{Hiroshima Synchrotron Radiation Center, Hiroshima University, 2-313 Kagamiyama, Higashi-Hiroshima 739-0046, Japan}
\author{Eike F. Schwier}
\affiliation{Hiroshima Synchrotron Radiation Center, Hiroshima University, 2-313 Kagamiyama, Higashi-Hiroshima 739-0046, Japan}
\author{Kenya Shimada}
\affiliation{Hiroshima Synchrotron Radiation Center, Hiroshima University, 2-313 Kagamiyama, Higashi-Hiroshima 739-0046, Japan}
\author{Lin Zhao}
\affiliation{Institute of Physics, Chinese Academy of Sciences, Beijing 100190, China}
\affiliation{School of Physical Sciences, University of Chinese Academy of Sciences, Beijing 100049, China}
\author{Kehui Wu}
\affiliation{Institute of Physics, Chinese Academy of Sciences, Beijing 100190, China}
\affiliation{School of Physical Sciences, University of Chinese Academy of Sciences, Beijing 100049, China}
\affiliation{Center of Materials Science and Optoelectronics Engineering, University of Chinese Academy of Sciences, Beijing 100049, China}
\affiliation{Songshan Lake Materials Laboratory, Dongguan, Guangdong 523808, China}
\author{Hongming Weng}
\affiliation{Institute of Physics, Chinese Academy of Sciences, Beijing 100190, China}
\author{Lan Chen}
\affiliation{Institute of Physics, Chinese Academy of Sciences, Beijing 100190, China}
\affiliation{School of Physical Sciences, University of Chinese Academy of Sciences, Beijing 100049, China}
\affiliation{Center of Materials Science and Optoelectronics Engineering, University of Chinese Academy of Sciences, Beijing 100049, China}
\affiliation{Songshan Lake Materials Laboratory, Dongguan, Guangdong 523808, China}
\author{X. J. Zhou}
\affiliation{Institute of Physics, Chinese Academy of Sciences, Beijing 100190, China}
\affiliation{School of Physical Sciences, University of Chinese Academy of Sciences, Beijing 100049, China}
\author{Zhijun Wang}
\email{wzj@iphy.ac.cn}
\affiliation{Institute of Physics, Chinese Academy of Sciences, Beijing 100190, China}
\affiliation{School of Physical Sciences, University of Chinese Academy of Sciences, Beijing 100049, China}
\author{Youguo Shi}
\email{ygshi@iphy.ac.cn}
\affiliation{Institute of Physics, Chinese Academy of Sciences, Beijing 100190, China}
\affiliation{School of Physical Sciences, University of Chinese Academy of Sciences, Beijing 100049, China}
\affiliation{Center of Materials Science and Optoelectronics Engineering, University of Chinese Academy of Sciences, Beijing 100049, China}
\affiliation{Songshan Lake Materials Laboratory, Dongguan, Guangdong 523808, China}
\author{Baojie Feng}
\email{bjfeng@iphy.ac.cn}
\affiliation{Institute of Physics, Chinese Academy of Sciences, Beijing 100190, China}
\affiliation{School of Physical Sciences, University of Chinese Academy of Sciences, Beijing 100049, China}

\date{\today}

\begin{abstract}
We report the synthesis of single-crystal NbC, a transition metal carbide with various unusual properties. Transport, magnetic susceptibility, and specific heat measurements demonstrate that NbC is a conventional superconductor with a superconducting transition temperature ($T_c$) of 11.5 K. Our theoretical calculations show that NbC is a type-II Dirac semimetal with strong Fermi surface nesting, which is supported by our ARPES measurement results. We also observed the superconducting gaps of NbC using angle-resolved photoemission spectroscopy (ARPES) and found some unconventional behaviors. These intriguing superconducting and topological properties, combined with the high corrosion resistance, make NbC an ideal platform for both fundamental research and device applications.
\end{abstract}

\maketitle

\section{INTRODUCTION}

Transition metal carbides (TMCs) are a large family of materials that have held industrial research interest for more than one hundred years. In TMCs, carbon atoms are incorporated into the interstitial sites of the parent metals and form strong metal-carbon bonds\cite{OyamaST,HwuHH2005,XiaoY2016}. This unique structure gives TMCs coexisting ionic, covalent, and metallic bonds, leading to metallic conductivity combined with excellent mechanical properties, including extreme hardness, high melting points, excellent corrosion resistance, and interesting catalysis properties\cite{OyamaST,HwuHH2005,XiaoY2016}. Therefore, the TMC family of materials has been widely used in a variety of areas, including cutting tools, energy storage devices, and supercapacitors.

One prototypical TMC material is niobium monocarbide (NbC), which crystallizes into a NaCl-type cubic structure. NbC has various outstanding aspects: its melting point is one of the highest among all solid materials, while its hardness (in the 9-10 range on the Mohs scale) is comparable to that of diamond\cite{Shabalin2019}. More interestingly, NbC is a superconductor and has a $T_c$ of approximately 11 K\cite{WellM1964,GengHX2007,ZouGF2011,JhaR2012}, which is higher than that of niobium. Recent theoretical studies have predicted Fermi surface nesting in NbC, which will enhance the electron-phonon interactions in the material\cite{BlackburnS2011,LiC2018}. Fermi surface nesting is a subject of intense interest in studies of high-temperature superconductivity and charge/spin density waves\cite{LeePA2006,WhangboMH1991,GiustinoF2017}. However, the biggest obstacle to the study of NbC is the challenge of synthesizing single-crystal samples. To date, only NbC polycrystals have been synthesized and this has strongly limited further experimental studies. Therefore, only a few experimental works on NbC and NbC$_x$ (0$<x<$1)\cite{Lindberg1987,Edamoto1989} have been performed, although the material was discovered more than half a century ago.

Recently, classification of electronic structures based on their topological properties has offered great opportunities to understand the behavior of solid-state materials\cite{YanB2017,ArmitageNP2018}. In particular, Dirac semimetals represent a novel class of quantum materials that is close to various exotic quantum phases, including topological insulators, Weyl semimetals, and topological superconductors. In Dirac semimetals, the bulk conduction bands and the valence bands touch linearly at discrete points to form three-dimensional Dirac cones and the degeneracy of the Dirac points is topologically protected. Additionally, superconducting Dirac semimetals may favor topological superconductivity, which hosts the long-sought Majorana fermions\cite{KobayashiS2015}. Considering the extraordinary properties in the TMCs mentioned above, the discovery of topological band structures in these TMCs would make them an excellent platform for novel quantum device fabrication.

In this work, we have successfully synthesized single-crystal NbC. Transport and magnetic susceptibility measurements confirmed that single-crystal NbC is a type-II superconductor that has a $T_c$ of 11.5 K. Angle-resolved photoemission spectroscopy (ARPES) measurements provided evidence of Fermi surface nesting, analogous to some iron-based superconductors and charge density wave systems. First-principles calculations showed that NbC hosts type-II Dirac cones near the Fermi level, which qualitatively agrees with our ARPES measurements. We also studied the superconducting gap using high-resolution laser-ARPES, and, interestingly, we found that the behavior of the gap deviates from that of conventional Bardeen-Cooper-Schrieffer (BCS) theory.

\section{METHODS}

The X-ray diffraction (XRD) measurements were carried out on a Bruker D8 Venture diffractometer at 293 K using Mo $\emph{K}\alpha$ radiation ($\lambda$ = 0.71073 {\AA}). The crystal structure was refined via the full-matrix least-squares method on \emph{F$^2$} using the SHELXL-2016/6 program. The electrical resistivity and the specific heat capacity were measured in a Physical Property Measurement System (PPMS, Quantum Design Inc.) using the standard dc four-probe technique and a thermal relaxation method, respectively. The magnetic properties of the material were measured using a Magnetic Properties Measurement System (MPMS-\uppercase\expandafter{\romannumeral3}, Quantum Design Inc.) under a fixed applied magnetic field of 20 Oe in the field cooling (FC) and zero field cooling (ZFC) modes. The isothermal magnetization (\emph{M-H}) was measured at several fixed temperatures by sweeping an applied field. The ARPES measurements were performed at beamline BL-1 of the Hiroshima Synchrotron Radiation Center\cite{IwasawaH2017}. The clean surfaces required for the ARPES measurements were obtained by cleaving the samples in situ along the (001) plane in an ultra-high vacuum chamber. The cleavage process and the measurements were performed at 30 K. The overall energy and angular resolutions were approximately 15 meV and 0.1$^{\circ}$, respectively. The superconducting gap was measured using a laser-based ARPES system equipped with a 6.994 eV laser and a time-of-flight electron energy analyzer\cite{ZhouXJ2018}. The energy resolution was approximately 1 meV and the lowest temperature achieved during the measurements was 1.9 K. The values of $k_z$ were determined using an inner potential of 10 eV.

Single-crystal NbC was grown by the solid-state reaction method using Co as a flux. We mixed high-purity Nb, C, and Co in an alumina crucible with a molar ratio of Nb:C:Co=1:1:9. The alumina crucible was then placed in an argon-filled furnace and heated to 1500 $^{\circ}$C. After a dwell time of 20 hours, the crucible was slowly cooled to 1300 $^{\circ}$C at a rate of 1$^{\circ}$C/h and then cooled naturally down to room temperature. Finally, the excess Co was removed by immersion in nitrohydrochloric acid for one day. The as-grown NbC single crystals are silver-grey in color with shiny rectangular faces.

We performed the first-principles calculations using the VASP package\cite{KresseG1996,KresseG1996prb} based on density functional theory (DFT) in combination with the projector augmented wave (PAW) method\cite{BlochlPE1994,KresseG1999}. The generalized gradient approximation (GGA) was used along with the Perdew-Burke-Ernzerhof (PBE) exchange-correlation functional\cite{PerdewJP1996}. The kinetic energy cutoff was set at 500 eV for the plane wave basis. A 12$\times$12$\times$12 $k$-mesh was used in a self-consistent process for Brillouin zone (BZ) sampling. The lattice and atomic parameters were obtained from our XRD data. The electronic structures with spin-orbit coupling (SOC) were then derived. The irreducible representations (irreps) were computed based on the high-symmetry points. The maximally localized Wannier function (MLWF) method was used to calculate the surface states\cite{MarzariN2012}. To fit well with the experimental data, the surface states are modified by adding a surface potential ($E$ = -1 eV) in the Green's function calculations.

\begin{figure}[htb]
\centering
\includegraphics[width=8cm]{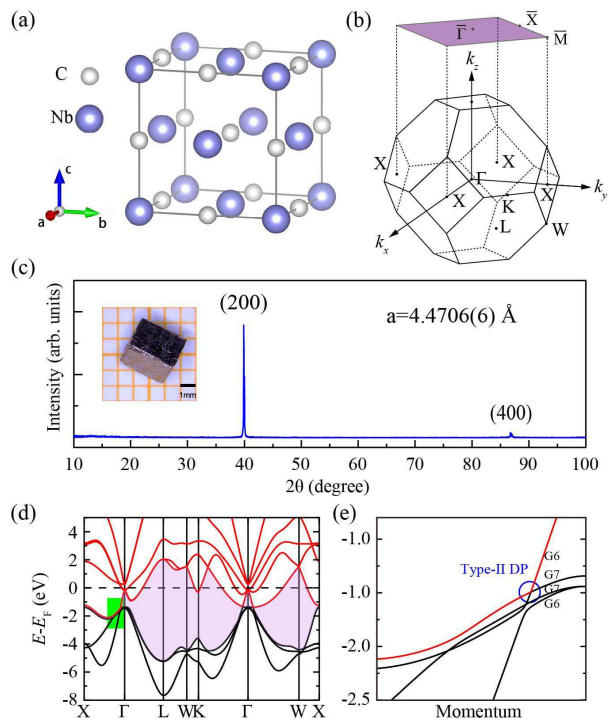}
\caption{Growth of single-crystal NbC and calculations on its electronic structures. (a) and (b) Schematic illustrations of the atomic structure and (001)-projected BZ of NbC, respectively. (c) X-ray diffraction pattern of the flat surface of the NbC single crystal. The inset shows a photograph of a typical NbC single crystal. (d) Calculated band structure of NbC, showing that a gap (gray shaded) exists between the black-colored bands and the red-colored bands, with the exception of a gapless Dirac point along $\Gamma$--X. (e) Magnified view of the band structure in the green shaded region of (d), which indicates the existence of a type-II Dirac cone. This cone is formed by irreducible representations of G6 and G7 under the double group of the $C_{4v}$ symmetry on the line.}
\end{figure}

\section{RESULTS AND DISCUSSIONS}

\begin{center}
\bf I. Transport, Magnetic Susceptibility, and Specific Heat Measurements
\end{center}

The growth methods and crystallographic parameters of single crystal NbC can be found in the Supplementary Materials \cite{SM}. NbC crystallized in a face-centered-cubic structure with a space group of {\it Fm\={3}m} (No. 225). The atomic structure and the Brillouin zone (BZ) of NbC are illustrated schematically in Fig. 1(a) and 1(b), respectively. Figure 1(c) shows an XRD spectrum that was measured on the (001) surface, which is a natural cleavage surface, and only the {\it h00} peaks were detected. The inset of Fig. 1(c) shows a photograph of a piece of NbC single crystal. The picture shows that the crystal is as large as several millimeters in size and has shiny rectangular faces, indicating the good crystallinity of the samples.

Figure 2(a) shows the magnetic susceptibility (4$\pi\chi$) of NbC as a function of temperature under zero field cooling (ZFC) and field cooling (FC) conditions at 20 Oe. The plots show that there is a sudden drop in the magnetic susceptibility at a critical temperature, which is a signature of the Meissner effect. Observation of the Meissner effect indicates occurrence of a superconducting transition and the fitted critical temperature $T_c$ is 11.5 K. The superconducting volume fraction that was calculated from the magnetic measurements exceeds 100\%, which originates from the demagnetization field caused by the NbC sample. The $M-H$ loops in Fig. 2(b) shows the existence of upper and lower critical fields, indicating that NbC is a type-II superconductor\cite{TinkhamM2004}. To provide further confirmation of the superconductivity in NbC, we measured the resistivity of NbC as a function of temperature, with results as shown in Fig. 2(c). As the temperature decreases from 300 K, the resistivity also decreases metallically until the temperature reaches 15 K; a sharp drop then occurs at an onset temperature of $T$ = 12.3 K. The resistivity finally decreases to zero at $T$ = 11.5 K, which agrees well with the $T_c$ value determined from the magnetic susceptibility measurements. In the presence of a magnetic field, the superconductivity is suppressed and $T_c$ shifts continuously toward a lower temperature as the magnetic field increases, as shown in the inset of Fig. 2(c).

\begin{figure}[htb]
\centering
\includegraphics[width=8.5cm]{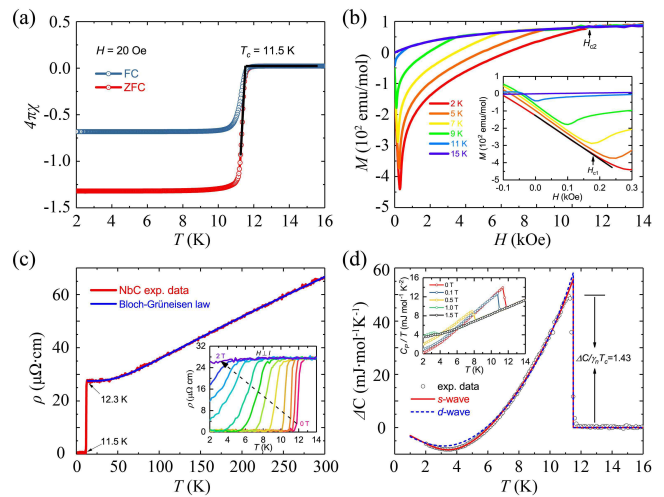}
\caption{Characterization of the basic physical properties of NbC. (a) Temperature dependence of 4$\pi$\emph{$\chi$} of NbC for an applied field $H$ = 20 Oe in the zero field cooling (ZFC) and field cooling (FC) modes. (b) Isothermal \emph{M-H} curves of NbC at various fixed temperatures. The inset shows an enlarged image of the \emph{M-H} curves below 300 Oe. (c) Temperature dependence of longitudinal resistivity $\rho$ of NbC. The inset shows the resistivity characteristics under different magnetic fields. (d) Experiment data for $\Delta$$C$ = $C$(0T)-$C$(2T) vs $T$, which were fitted using $s$-wave and $d$-wave gap functions. The inset shows the $C_P$/$T$ versus $T$ curves under various applied magnetic fields.}
\end{figure}

The combination of the magnetic susceptibility and transport measurements confirmed the occurrence of the superconducting transition in NbC unambiguously. We then fitted the electron-phonon coupling (EPC) constant $\lambda_{ep}$ using the inverted McMillan equation\cite{MiMillan1968}:

\begin{equation}
\lambda_{ep}=\frac{1.04+\mu^*\ln(\frac{\theta_D}{1.45T_c})}{(1-0.62\mu^*)\ln(\frac{\theta_D}{1.45T_c})-1.04}
\end{equation}

where $\theta_D$=321.6 K is the Debye temperature, which can be fitted based on the resistivity data measured above 15 K using the Bloch-Gr\"{u}neisen law; $\mu^*$ represents the repulsively screened Coulomb part, which is set at 0.13 for the transition metal element; and $T_c$ is the superconducting transition temperature of 11.5 K. The fitted EPC constant $\lambda_{ep}$ is 0.848, which is slightly higher than the theoretically calculated value of 0.682 from a previous report\cite{BlackburnS2011}.

\begin{figure*}[htb]
\centering
\includegraphics[width=17 cm]{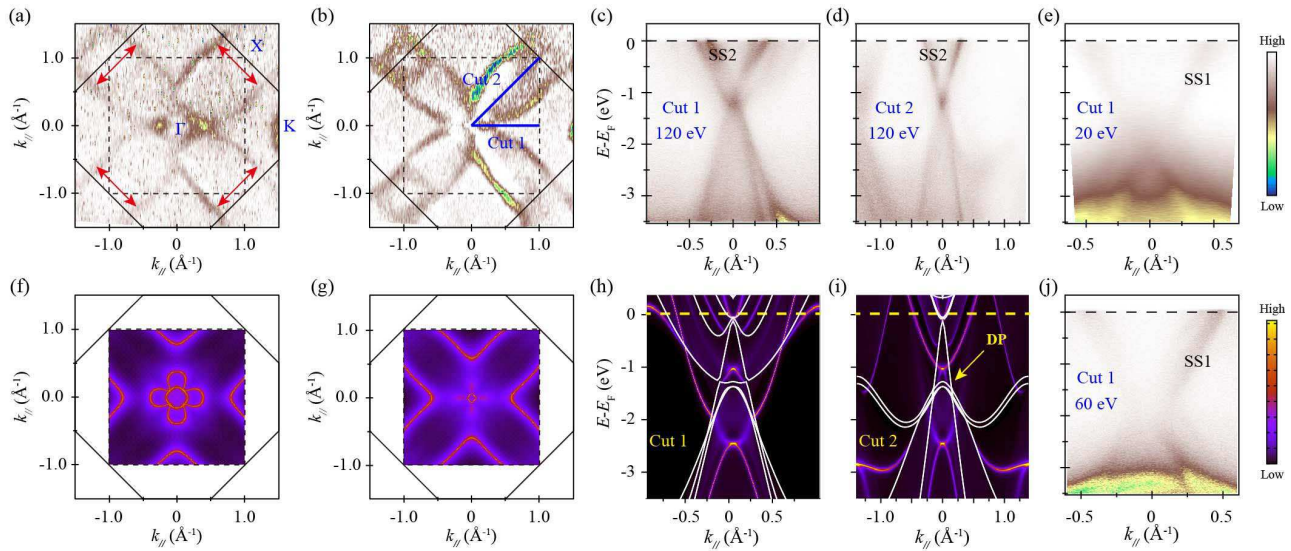}
\caption{Synchrotron ARPES measurements and theoretical calculations of NbC. (a) ARPES intensity map of the Fermi surface of NbC measured using 120 eV photons. Solid and dashed lines indicate the bulk and surface BZs, respectively. Red arrows indicate the possible nesting vectors. (b) ARPES intensity map of the constant energy contour (CEC) at $E_B$=0.6 eV measured using 120 eV photons. (c,d) ARPES intensity plots of band structures along $\Gamma$-K and $\Gamma$-X measured using 120 eV photons. (e,j) ARPES intensity plots of band structures along $\Gamma$-K measured using 20 eV and 60 eV photons, respectively. (f,g) Calculated bulk bands on the Fermi surface and CEC at $E_B$=0.6 eV. (h,i) Purple-yellow colored images are semi-infinite calculation results using the Green's function method along directions that are parallel to $\Gamma$--K and $\Gamma$--X, respectively. The surface states can be determined as sharp yellow lines. White lines are the calculated bulk bands along $\Gamma$--K and $\Gamma$--X. ``SS1'' and ``SS2'' indicate two surface states of NbC.}
\end{figure*}

In addition to the magnetic susceptibility and the resistivity, the heat capacity also shows a jump at $T_c$, and as expected, this jump tends to be suppressed as the magnetic field increases, as shown in Fig. 2(d). When the normal state data $C$(2T) are subtracted from the zero-field heat capacity $C$(0T), the resulting $\Delta$$C$=$C$(0T)-$C$(2T) removes the phonon contribution and the nonsuperconducting part to yield $\Delta$$C$ = $C_{es}$ - $\gamma_n$$T$, where $C_{es}$ is the contribution of the Cooper pairs and $\gamma_n$ is the normal state Sommerfeld coefficient of the superconducting part\cite{TaylorOJ2007,XuX2013,NiuCQ2013}. We fitted $\Delta$$C$ using $s$ and $d$ wave models and the fitting results are as shown in Fig. 2(d). We found that the $s$-wave model can reproduce our experimental data well, while the $d$-wave model showed significant deviation from the data at low temperatures. From the $s$-wave fitting, we obtained $\gamma_n$ = 3.01 mJ mol$^{-1}$ K$^{-2}$ and $T_c$ = 11.5 K. Consequently, the estimated superconducting gap $\Delta$(0) is 1.85 meV and the gap ratio 2$\Delta$(0)/$k_B$$T_c$ is approximately 3.73, which is quite close to the value obtained from the BCS theory (3.52). The entropy-conserving construction at $T_c$ gives $\Delta$C/$\gamma_n$$T_c$ = 1.43, which is equal to the weak coupling BCS value.

\begin{center}
\bf II. Electronic Band Structure Calculations
\end{center}

Having confirmed the superconductivity of single-crystal NbC, we move on to study the electronic structures of NbC. Figure 3(a) shows an ARPES intensity map of the Fermi surface of the (001) surface measured at a photon energy of 120 eV, which approximately corresponds to the $k_z$=0 plane of the bulk BZ. Flower-like bands with four lobes were observed near the $\Gamma$ point, along with several straight bands that lay parallel to $\Gamma$--X. The ARPES intensities of the Fermi surface, in particular the straight bands, were in qualitative agreement with our calculated bulk bands, as shown in Fig. 3(f). The parallel bands at the Fermi level indicate the existence of Fermi surface nesting, as indicated by the red arrows shown in Fig. 3(a). The nesting vector obtained from our experiments is approximately 0.85 \AA$^{-1}$. Previous calculations indicate that Fermi surface nesting in NbC will lead to the emergence of the Kohn anomaly and will enhance the electron-phonon coupling in NbC\cite{NoffsingerJ2008,BlackburnS2011}. According to the BCS theory, stronger electron-phonon coupling will increase the critical temperatures of superconductors. In addition, such nesting condition is absent in isostructural compound TiC\cite{IsaevEI2007}, which indicates that Fermi surface nesting may be a requisite condition for the high $T_c$ in the NbC family of materials. With the increase of binding energies, the parallel bands move closer together, as shown in Fig. 3(b) and 3(g).

Next, we discuss the topological properties of NbC based on our calculated results. The calculated electronic band structure of NbC is shown in Fig. 1(d), which agrees well with previous calculations\cite{BlackburnS2011,LiC2018,AmriouT2003}. To do analysis of the elementary band representations (EBRs) in terms of the theory of topological quantum chemistry\cite{BradlynB2017}, we have obtained irreducible representations at the maximal high-symmetry points\cite{VergnioryM2019}, as presented in Table S4 of the Supplementary Materials \cite{SM}. The results of our EBR analysis show that NbC is a symmetry-enforced semimetal with six (black-colored) occupied bands, which means that NbC must have some symmetry-protected nodes along high-symmetry momentum lines (see more on the band-inverted feature in the Supplementary Materials \cite{SM}). Detailed calculations show that there are six Dirac points along these high-symmetry lines (Fig. 1(e) or the green shaded region of Fig. 1(d)). On the $\Gamma$--X line, the Dirac point is formed by irreps G6 and G7 of the $C_{4v}$ symmetry. It can be found that the Dirac cone is strongly tilted, which indicates that it is a type-II Dirac cone.

\begin{center}
\bf III. ARPES Measurements
\end{center}

\begin{figure}[htb]
\centering
\includegraphics[width=8cm]{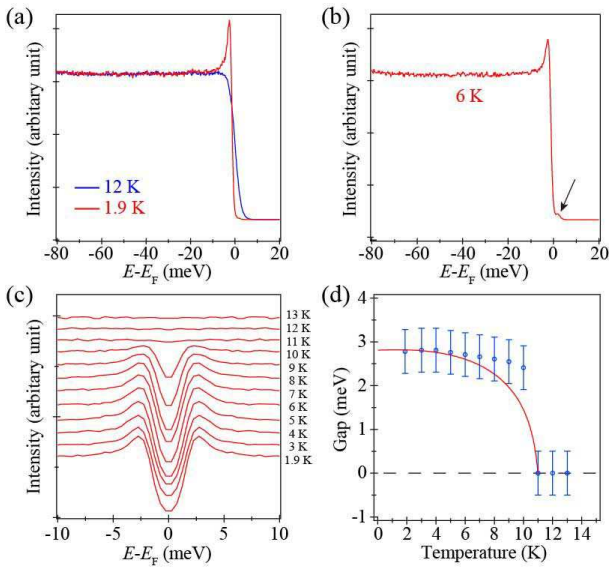}
\caption{Laser-ARPES measurements of the superconducting gap of NbC. (a) Energy distribution curves (EDCs) of NbC near the Fermi level, as measured at 1.9 K (red) and 12 K (blue). The pronounced peak observed at 1.9 K corresponds to the Bogoliubov quasiparticle coherence peak. (b) EDC measured at 6 K. The black arrow indicates the coherence peak that lies above the Fermi level. (c) Symmetrized EDCs of NbC at different temperatures. (d) Temperature dependence of the superconducting gap. The gap was fitted using the symmetrized EDCs given in (c). The red curve was calculated based on BCS theory. The EDCs were integrated in a momentum range of -0.12 {\AA}$^{-1}$ $<$ $k_x$ $<$ 0.12 {\AA}$^{-1}$, -0.12 {\AA}$^{-1}$ $<$ $k_y$ $<$ 0.12 {\AA}$^{-1}$. The incident photon energy was 6.994 eV.}
\end{figure}

The ARPES intensity maps along $\Gamma$--K and $\Gamma$--X are shown in Fig. 3(c) and 3(d), respectively. The Dirac point cannot be observed clearly in our ARPES results because of the intrinsic $k_z$ broadening of the bulk bands. To better understand the experimental results, we calculated the surface states of NbC and the results are shown in Fig. 3(h) and 3(i). These surface states originate from the band inversion induced by the strong crystal field. The bulk bands are superimposed in order to facilitate comparison. A detailed comparison with the calculation results shows qualitative agreement between experimental and theoretical results. Along $\Gamma$--K and $\Gamma$--X, the experimental band structures at 120 eV contain contributions of both bulk bands and surface states. The surface states can be better resolved with lower photon energies, as shown in Fig. 3(e) and 3(j).

Finally, we studied the superconducting gap using high-resolution laser-ARPES. At low photon energy, the bands of NbC have small photoemission cross-section and are thus almost invisible in our measurements (see Fig. S5). Therefore, we presented the integrated energy distribution curves (EDCs), as shown in Fig. 4(a). At 1.9 K, the sharp peak that occurs slightly below the Fermi level originates from the coherence of the Bogoliubov quasiparticles. The peak intensity decreases with increasing temperature and the peak finally disappears when the temperature approaches $T_c$. The assignment of the Bogoliubov quasiparticle peak is further confirmed by the emergence of the coherence peak above the Fermi level at a slightly higher temperature, as indicated by the black arrow in Fig. 4(b). At 1.9 K, the temperature is too low and the Fermi-Dirac distribution is quite sharp. In this case, the density of states above $E\rm_F$ is negligible and the coherence peak above $E_F$ cannot therefore be detected using conventional ARPES. At a higher temperature, e.g., at 6 K, the Fermi-Dirac distribution becomes broader and this leads to the emergence of the coherence peak above $E\rm_F$.

To provide a better visualization of the superconducting gap, we symmetrized the EDCs at the various temperatures, as shown in Fig. 4(c). Figure 4(d) shows the fitted gap size as a function of temperature. When the temperature falls below 4 K, the gap is approximately 2.8 meV and shows little variation with temperature. As the temperature increases, the gap slowly decreases to 2.4 meV at 10 K before dropping suddenly to zero at $T_c$. This behavior obviously deviates from that of BCS theory, as indicated by the red curve. This sudden closure of the superconducting gap is reminiscent of some cuprate\cite{Norman1998} and iron-based superconductors\cite{DingH2008,ZhaoL2008}. Furthermore, the calculated (2$\Delta_0$)/($k_B$$T_c$) has a value of approximately 5.7, which is much larger than the corresponding value for conventional BCS-type superconductors ($\sim$3.5). This value is also larger than the value that was obtained from our specific heat data. One possible reason for this difference is that the photoemission signal originates from the topmost atomic layers of NbC, which may have properties that differ from those of the bulk.

\begin{center}
\bf IV. Discussion
\end{center}

The coexistence of superconductivity and the bulk Dirac cone in NbC provides opportunities to explore the novel properties in the material\cite{SakanoM2015,LiY2019}. Recent theoretical works have predicted that the coexistence of superconductivity and bulk Dirac cones can cause three-dimensional topological superconductivity\cite{KobayashiS2015}. In NbC, however, the Dirac points are located at 1.5 eV below the Fermi level; the realization of three-dimensional topological superconductivity requires the tuning of the Dirac bands to the proximity of the Fermi level. On the other hand, Dirac semimetal hosts topological surface states that connect the Dirac points on certain side surfaces. When an s-wave superconductor is in close proximity to topological surface states, topological superconducting states can be expected at the interface\cite{FuL2008,ZhangP2018}. In NbC, however, the coexistence of the bulk s-wave superconductivity and topological surface states might induce topological superconductivity at the NbC surface itself. In addition, our laser-ARPES results for the gap behaviors suggest that the superconducting properties at the surface region differ from those in the bulk, although the values of $T_c$ at the surface and the bulk are the same, and this seems to be a signature of bulk-induced unconventional superconductivity at the surface. Similar unconventional gap features at a material surface have been observed by scanning tunneling spectroscopy in the topological superconductors $\beta$-Bi$_2$Pb\cite{LvYF2017} and MoTe$_{2-x}$S$_x$\cite{LiY2018}.

\section{CONCLUSIONS}

Our results demonstrate that superconductivity and topological band structures can coexist in stoichiometric single-crystal NbC and, more interestingly, the superconducting properties deviate from those given by BCS theory at the surface region. In addition, NbC shows strong Fermi surface nesting; this nesting contributes to the strong electron-phonon interactions and could ultimately enhance the superconducting order of the material. These intriguing properties mean that NbC is a material of fundamental interest for the exploration of exotic properties such as topological superconductivity and the Majorana bound states. Additionally, as a member of the TMC family of materials, NbC is highly stable: it remains intact in nitrohydrochloric acid, sodium hydroxide solution and organic solvents, and its melting point and hardness are among the highest of all known materials. This excellent stability could enable the operation of NbC-based devices under extreme conditions.

{\it Note added.} During the preparation of a revised manuscript, we became aware of an independent work on the characterization of polycrystalline NbC and TaC and prediction of topological properties \cite{TangS2020}.

\begin{acknowledgments}
This work was supported by the National Key Research and Development Program of China (Grant No. 2017YFA0302901, No. 2016YFA0300604, and No. 2016YFA0300300), the MOST of China (Grant No. 2018YFE0202700), the National Natural Science Foundation of China (Grants No. 11761141013, No. 11888101, No. 11974391, and No. 11774399), the Beijing Natural Science Foundation (Grant No. Z180007 and No. Z180008), the Strategic Priority Research Program of the Chinese Academy of Sciences (Grants No. XDB33030100 No. XDB30000000), and the K. C. Wong Education Foundation (GJTD-2018-01). The synchrotron ARPES measurements were performed with the approval of the Proposal Assessing Committee of Hiroshima Synchrotron Radiation Center (Proposal Numbers: 18BG010 and 18AG017).
\end{acknowledgments}

\clearpage


\begin{thebibliography}{99}
{
\bibitem{OyamaST} S. T. Oyama, The chemistry of transition metal carbides and nitrides. (Springer 1996).

\bibitem{HwuHH2005} H. H. Hwu and J. G. Chen, Surface chemistry of transition metal carbides. Chem. Rev. {\bf 105}, 185 (2005).

\bibitem{XiaoY2016} Y. Xiao, J.-Y. Hwang, and Y.-K. Sun, Transition metal carbide-based materials: synthesis and applications in electrochemical energy storage. J. Mater. Chem. A. {\bf 4}, 10379 (2016).

\bibitem{Shabalin2019} I. L. Shabalin, Ultra-high temperature materials II. (Springer Nature B.V. 2019)

\bibitem{WellM1964} M. Wells, M. Pickus, K. Kennedy, and V. Zackay, Superconductivity of solid solutions of TaC and NbC. Phys. Rev. Lett. {\bf 12}, 536-537 (1964).

\bibitem{GengHX2007} H. X. Geng, G. C. Che, W. W. Huang, S. L. Jia, H. Chen, and Z. X. Zhao, Structural, morphological, C content and TC changes of the NbC superconductor prepared by Nb powder and carbon nanotubes. Supercond. Sci. Technol. {\bf 20}, 211-215 (2007).

\bibitem{ZouGF2011} G. F. Zou, H. M. Luo, S. Baily, Y. Y. Zhang, N. F. Haberkorn, J. Xiong, E. Bauer, T. M. McCleskey, A. K. Burrell, L. Civale, Y. T. Zhu, J. L. MacManus-Driscoll, and Q. X. Jia, Highly aligned carbon nanotube forests coated by superconducting NbC. Nat. Commun. {\bf 2}, 428 (2011).

\bibitem{JhaR2012} R. Jha and V. P. S. Awana, Vacuum encapsulated synthesis of 11.5 K NbC superconductor. J. Supercond. Nov. Magn. {\bf 25}, 1421 (2012).

\bibitem{BlackburnS2011} S. Blackburn, Michel C\^{o}t\'{e}, S. G. Louie, and M. L. Cohen, Enhanced electron-phonon coupling near the lattice instability of superconducting NbC$_{1-x}$N$_x$ from density-functional calculations. Phys. Rev. B {\bf 84}, 104506 (2011).

\bibitem{LiC2018} C. Li, N. K. Ravichandran, L. Lindsay, and D. Broido, Fermi surface nesting and phonon frequency gap drive anomalous thermal transport. Phys. Rev. Lett. {\bf 121}, 175901 (2018).

\bibitem{LeePA2006} P. A. Lee, N. Nagaosa, and X.-G. Wen, Doping a Mott insulator: Physics of high-temperature superconductivity. Rev. Mod. Phys. {\bf 78}, 17 (2006).

\bibitem{WhangboMH1991} M.-H. Whangbo, E. Canadell, P. Foury, and J.-P. Pouget, Hidden Fermi surface nesting and charge density wave instability in low-dimensional metals. Science {\bf 252}, 96 (1991).

\bibitem{GiustinoF2017} F. Giustino, Electron-phonon interactions from first principles. Rev. Mod. Phys. {\bf 89}, 015003 (2017).

\bibitem{Lindberg1987} P. A. P. Lindberg, L. I. Johansson, J. B. Lindstr\"{o}m, and P. E. S. Persson, Phys. Rev. B {\bf 36}, 6343 (1987).

\bibitem{Edamoto1989} K. Edamoto, S. Maehama, and E. Miyazaki, Electronic structure of a NbC$_{0.9}$(100) surface: angle-resolved photoemission study. Phys. Rev. B {\bf 39}, 7461 (1989).

\bibitem{YanB2017} B. Yan and C. Felser, Topological materials: Weyl semimetals. Annu. Rev. Condens. Matter Phys. {\bf 8}, 337 (2017).

\bibitem{ArmitageNP2018} N. P. Armitage, E. J. Mele, and A. Vishwanath, Weyl and Dirac semimetals in three-dimensional solids. Rev. Mod. Phys. {\bf 90}, 015001 (2018).

%calculations

\bibitem{KobayashiS2015} S. Kobayashi and M. Sato, Topological Superconductivity in Dirac Semimetals. Phys. Rev. Lett. {\bf 115}, 187001 (2015).

\bibitem{IwasawaH2017} H. Iwasawa, K. Shimada, E. F. Schwier, M. Zheng, Y. Kojima, H. Hayashi, J. Jiang, M. Higashiguchi, Y. Aiura, H. Namatamea, and M. Taniguchi, Rotatable high-resolution ARPES system for tunable linear-polarization geometry. J. Synchrotron Radiat. {\bf 24}, 836 (2017).

\bibitem{ZhouXJ2018} X. J. Zhou, S. He, G. Liu, L. Zhao, L. Yu, and W. Zhang, New developments in laser-based photoemission spectroscopy and its scientific applications: a key issues review. Rep. Prog. Phys. {\bf 81}, 062101 (2018).

\bibitem{KresseG1996} G. Kresse and Furthm\"{u}ller, J. Efficiency of ab-initio total energy calculations for metals and semiconductors using a plane-wave basis set. Comput. Mater. Sci. {\bf 6}, 15 (1996).

\bibitem{KresseG1996prb} G. Kresse and J. Furthm\"{u}ller, Efficient iterative schemes for ab initio total-energy calculations using a plane-wave basis set. Phys. Rev. B {\bf 54}, 11169 (1996).

\bibitem{BlochlPE1994} Bl\"{o}chl and P. E. Projector augmented-wave method. Phys. Rev. B 50, 17963 (1994).

\bibitem{KresseG1999} G. Kresse and D. Joubert, From ultrasoft pseudopotentials to the projector augmented-wave method. Phys. Rev. B {\bf 59}, 1758 (1999).

\bibitem{PerdewJP1996} J. P. Perdew, K. Burke, and M. Ernzerhof, Phys. Rev. Lett. {\bf 77}, 3865 (1996).

\bibitem{MarzariN2012} N. Marzari, A. A. Mostofi, J. R. Yates, I. Souza, and D. Vanderbilt, Rev. Mod. Phys. {\bf 84}, 1419 (2012).

\bibitem{SM} See Supplemental Material at [URL will be inserted by publisher] for additional experimental and theoretical data, which includes Ref. \cite{WerthamerNR1966,WeiW2016,TothL1971,StrocovVN2003,KondoT2009,HuangJ2019}.

%SM
\bibitem{WerthamerNR1966} N. R. Werthamer, E. Helfand, and P. C. Hohenberg, Temperature and Purity Dependence of the Superconducting Critical Field, H$_{c2}$. III. Electron Spin and Spin-Orbit Effects. Phys. Rev. {\bf 147}, 295 (1966).

\bibitem{WeiW2016} W. Wei, G. J. Zhao, D. R. Kim, C. Jin, J. L. Zhang, L. Ling, L. Zhang, H. Du, T. Y. Chen, J. Zang, M. Tian, C. L. Chien, and Y. Zhang, Rh$_2$Mo$_3$N: Noncentrosymmetric s-wave superconductor. Phys. Rev. B {\bf 94}, 104503 (2016).

\bibitem{TothL1971} L. Toth, Transition Metal Carbides and Nitrides 1st Edition. Elsevier (1971).

\bibitem{StrocovVN2003} V. N. Strocov, Intrinsic accuracy in 3-dimensional photoemission band mapping. J. Electron Spectrosc. Relat. Phenom. {\bf 130}, 65 (2003).

\bibitem{KondoT2009} T. Kondo, R. Khasanov, T. Takeuchi, J. Schmalian, and A. Kaminski, Competition between the pseudogap and superconductivity in the high-T$_c$ copper oxides. Nature {\bf 457}, 296 (2009).

\bibitem{HuangJ2019} J. Huang, L. Zhao, C. Li, Q. Gao, J. Liu, Y. Hu, Y. Xu, Y. Cai, D. Wu, Y. Ding, C. Hu, H. Zhou, X. Dong, G. Liu, Q. Wang, S. Zhang, Z. Wang, F. Zhang, F. Yang, Q. Peng, Z. Xu, C. Chen, and X. J. Zhou, Emergence of superconductivity from fully incoherent normal state in an iron-based superconductor (Ba$_{0.6}$K$_{0.4}$)Fe$_2$As$_2$. Sci. Bull. {\bf 64}, 11 (2019).
%SM

\bibitem{TinkhamM2004} M. Tinkham, Introduction to superconductivity. (Dover Publications, 2004).

\bibitem{MiMillan1968} W. L. McMillan, Transition temperature of strong-coupled superconductors. Phy. Rev. {\bf 167}, 331 (1968)

\bibitem{TaylorOJ2007} O. J. Taylor, A. Carrington, and J. A. Schlueter, Specific-heat measurements of the gap structure of the organic superconductors $\kappa$-(ET)$_2$Cu[N(CN)$_2$]Br and $\kappa$-(ET)$_2$Cu(NCS)$_2$. Phys. Rev. Lett. {\bf 99}, 057001 (2007).

\bibitem{XuX2013} X. Xu, B. Chen, W. H. Jiao, B. Chen, C. Q. Niu, Y. K. Li, J. H. Yang, A. F. Bangura, Q. L. Ye, C. Cao, J. H. Dai, G. Cao, and N. E. Hussey, Evidence for two energy gaps and Fermi liquid behavior in the SrPt$_2$As$_2$ superconductor. Phys. Rev. B {\bf 87}, 224507 (2013).

\bibitem{NiuCQ2013} C. Q. Niu, J. H. Yang, Y. K. Li, N. Zhou, J. Chen, L. L. Jiang, B. Chen, X. X. Yang, Chao Cao, J. Dai, and X. Xu, Effect of selenium doping on the superconductivity of Nb$_2$Pd(S$_{1-x}$Se$_x$)$_5$. Phys. Rev. B {\bf 88}, 104507 (2013).

\bibitem{NoffsingerJ2008} J. Noffsinger, F. Giustino, S. G. Louie, and M. L. Cohen, First-principles study of superconductivity and Fermi-surface nesting in ultrahard transition metal carbides. Phys. Rev. B {\bf 77}, 180507(R) (2008).

\bibitem{VergnioryM2019} M. Vergniory, L. Elcoro, C. Felser, N. Regnault, B. A. Bernevig, and Z. Wang, A complete catalogue of high-quality topological materials. Nature {\bf 566}, 480 (2019).

\bibitem{IsaevEI2007} E. I. Isaev, S. I. Simak, I. A. Abrikosov, R. Ahuja, Yu. Kh. Vekilov, M. I. Katsnelson, A. I. Lichtenstein, and B. Johansson, Phonon related properties of transition metals, their carbides, and nitrides: A first-principles study. J. Appl. Phys. {\bf 101}, 123519 (2007).

\bibitem{AmriouT2003} T. Amriou, B. Bouhafs, H. Aourag, B. Khelifa, S. Bresson, and C. Mathieu, FP-LAPW investigations of electronic structure and bonding mechanism of NbC and NbN compounds. Phys. B {\bf 325}, 46 (2003).

\bibitem{BradlynB2017} B. Bradlyn, L. Elcoro, J. Cano, M. Vergniory, Z. Wang, C. Felser, M. I. Aroyo, and B. A. Bernevig, Topological quantum chemistry. Nature {\bf 547}, 298 (2017).

\bibitem{Norman1998} M. R. Norman, H. Ding, M. Randeria, J. C. Campuzano, T. Yokoya, T. Takeuchi, T. Takahashi, T. Mochiku, K. Kadowaki, P. Guptasarma, and D. G. Hinks, Destruction of the Fermi surface in underdoped high-T$c$ superconductors. Nature {\bf 392}, 157 (1998).

\bibitem{DingH2008} H. Ding, P. Richard, K. Nakayama, K. Sugawara, T. Arakane, Y. Sekiba, A. Takayama, S. Souma, T. Sato, T. Takahashi, Z. Wang, X. Dai, Z. Fang, G. F. Chen, J. L. Luo and N. L. Wang, Observation of Fermi-surface-dependent nodeless superconducting gaps in Ba$_{0.6}$K$_{0.4}$Fe$_2$As$_2$, EPL {\bf 83}, 47001 (2008).

\bibitem{ZhaoL2008} L. Zhao, H.-Y. Liu, W.-T. Zhang, J.-Q. Meng, X.-W. Jia, G.-D. Liu, X.-L. Dong, G.-F. Chen, J.-L. Luo, N.-L. Wang, W. Lu, G.-L. Wang, Y. Zhou, Y. Zhu, X.-Y. Wang, Z.-Y. Xu, C.-T. Chen, and X. J. Zhou, Multiple nodeless superconducting gaps in (Ba$_{0.6}$K$_{0.4}$)Fe$_2$As$_2$ superconductor from angle-resolved photoemission spectroscopy. {\it Chin. Phys. Lett.} {\bf 25}, 4402 (2008).

\bibitem{SakanoM2015} M. Sakano, K. Okuwa, M. Kanou, H. Sanjo, T. Okuda, T. Sasagawa, and K. Ishizaka, Topologically protected surface states in a centrosymmetric superconductor $\beta$-PdBi$_2$. Nat. Commun. {\bf 6}, 8595 (2015).

\bibitem{LiY2019} Y. Li, X. Xu, M.-H. Lee, M.-W. Chu, and C. L. Chien, Observation of half-quantum flux in the unconventional superconductor $\beta$-Bi$_2$Pd. Science {\bf 366}, 238 (2019).

\bibitem{FuL2008} L. Fu and C. L. Kane, Superconducting proximity effect and Majorana fermions at the surface of a topological insulator. Phys. Rev. Lett. {\bf 100}, 096407 (2008).

\bibitem{ZhangP2018} P. Zhang, Z. Wang, X. Wu, K. Yaji, Y. Ishida, Y. Kohama, G. Dai, Y. Sun, C. Bareille, K. Kuroda, T. Kondo, K. Okazaki, K. Kindo, X. Wang, C. Jin, J. Hu, R. Thomale, K. Sumida, S. Wu, K. Miyamoto, T. Okuda, H. Ding, G. D. Gu, T. Tamegai, T. Kawakami, M. Sato, and S. Shin, Multiple topological states in iron-based superconductors. Nat. Phys. {\bf 15}, 41 (2018).

\bibitem{LvYF2017} Y.-F. Lv, W.-L. Wang, Y.-M. Zhang, H. Ding, W. Li, L. Wang, K. He, C.-L. Song, X.-C. Ma, and Q.-K. Xue, Experimental signature of topological superconductivity and Majorana zero modes on $\beta$-Bi$_2$Pd thin films. Sci. Bull. {\bf 62}, 852 (2017).

\bibitem{LiY2018} Y. Li, Q. Gu, C. Chen, J. Zhang, Q. Liu, X. Hu, J. Liu, Y. Liu, L. Ling, M. Tian, Y. Wang, N. Samarth, S. Li, T. Zhang, J. Feng, and J. Wang, Nontrivial superconductivity in topological MoTe$_{2-x}$S$_x$ crystals. Proc. Natl. Acad. Sci. U.S.A {\bf 115}, 9503 (2018).

\bibitem{TangS2020} T. Shang, J. Z. Zhao, D. J. Gawryluk, M. Shi, M. Medarde, E. Pomjakushina, and T. Shiroka, Superconductivity and topological aspects of the rocksalt carbides NbC and TaC. Phys. Rev. B {\bf 101}, 214518 (2020).

}
\end{thebibliography}
\end{document}